# Laguerre-Gaussian mode sorter

Nicolas K. Fontaine[1,†], Roland Ryf[1], Haoshuo Chen[1], David T. Neilson[1], Kwangwoong Kim[1] and Joel Carpenter[2,†,*]

**Light's spatial properties represent an infinite state space, making it attractive for applications requiring high dimensionality, such as quantum mechanics and classical telecommunications, but also inherently spatial applications such as imaging and sensing. However, there is no demultiplexing device in the spatial domain comparable to a grating or calcite for the wavelength and polarisation domains respectively. Specifically, a simple device capable of splitting a finite beam into a large number of discrete spatially separated spots each containing a single orthogonal spatial component. We demonstrate a device capable of decomposing a beam into a Cartesian grid of identical Gaussian spots each containing a single Laguerre-Gaussian component. This is the first device capable of decomposing the azimuthal and radial components simultaneously, and is based on a single spatial light modulator and mirror. We demonstrate over 210 spatial components, meaning it is also the highest dimensionality mode multiplexer of any kind.**

In order for a given property of light to be exploited for a particular application, it is first necessary to have components capable of discriminating in the basis of that particular property. For example, in the wavelength domain there are many dispersive elements such as gratings or prisms which can take a beam of light and map its spectral components to completely separated spatial positions. From there, each component can be analysed and/or manipulated. Similarly, in the polarisation domain, there are many types of polarising beamsplitters capable of splitting a beam into two spatially separated beams each containing an orthogonal polarisation component. These are examples of fundamental operations used routinely throughout optical physics with origins dating back centuries. The ability to perform these operations is largely taken for granted by researchers and optical system designers, and is often a trivial task which can be implemented using any of a large variety of available components.

In the general case, separating a beam into its orthogonal spatial components is inherently more difficult. However some spatial basis transformations are far easier than others. For example, the ability of a single positive lens, or infinite free-space propagation, to perform the Fourier transform is a simple example of a spatial mode decomposition of great practical importance and practical simplicity. A lens transforms real-space to angle-space ($k$-space), with a single position $(x,y)$ at the focus of the lens being transformed into a tilted plane-wave $(k_x,k_y)$ at the focus on the other side[1]. However the continuous and unbounded nature of a true plane-wave is such that it is not always a desirable basis.

Arguably the next most widely used spatial basis after Fourier is Laguerre-Gaussian (LG)[1]. Yet despite being a fundamental spatial basis in beam optics, unlike Fourier decomposition, which can be performed easily using a lens, there has previously been no analogous optical element for LG decomposition. The LG basis has several properties which make it an attractive basis throughout beam optics, fibre optics and resonators. LG (and the degenerate Hermite-Gaussian (HG)) basis are solutions of the paraxial wave equation, as well as eigenmodes of the Fourier transform and parabolic refractive index profiles such as a graded index fibre or GRIN lenses. LG modes are also circularly symmetric and directly related to the orbital angular momentum (OAM) of the photon[2], [3]. As a 2D orthogonal set, each mode in the LG basis is denoted by two indices. The radial index[4], [5], $\rho$, and the azimuthal index[2], [3], $l$, representing the quantised orbital angular momentum per photon.

Various mode (de)multiplexers, also called 'mode sorters', have been developed over the years which have implemented some limited ability to decompose a beam into its orthogonal LG components. Most approaches are able to decompose in only one-dimension, typically the azimuthal[6]–[8] component, or in the past year, the 'forgotten'[9] radial component[10], [11], but not simultaneous sorting of both, or in a non-orthogonal fashion[12]. Although successive stages for azimuthal and radial components could be cascaded together to address both indices, this scales poorly to large mode counts. Packing the highest number of degrees of freedom into a single photon or beam cross section is often one of the primary motivations for the use of LG beams in quantum optics[13]–[20] and classical communications[21]–[27]. Yet sampling a 2D function like a beam's transverse field, along just a single 1D axis means the vast majority of the possible degrees of freedom will always be missing. To completely describe any arbitrary 2D transverse field in the LG basis, and fully exploit the high dimensionality such beams provide, both azimuthal and radial components are required.

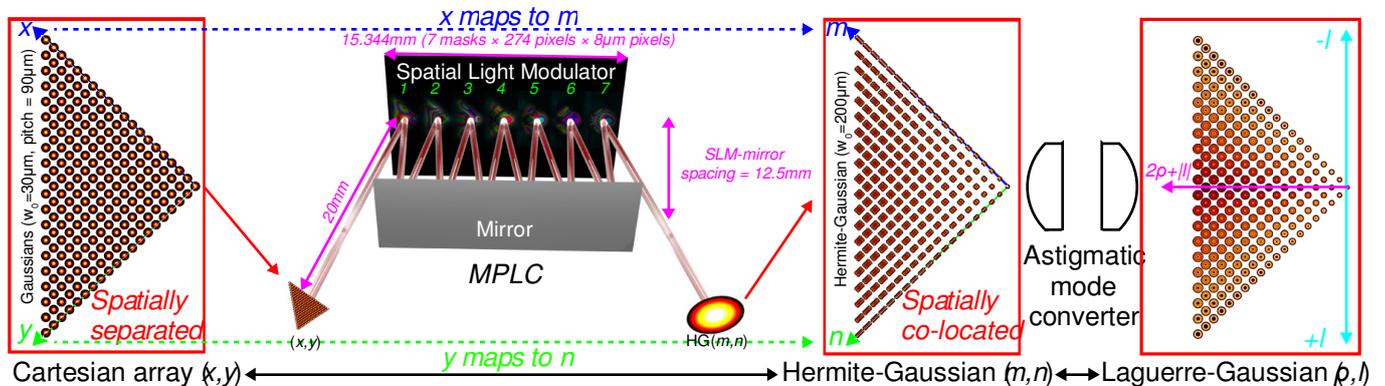

Figure 1 - Laguerre-Gaussian mode sorter based on multi-plane light conversion (MPLC). A Cartesian grid of Gaussian spots (MFD = 60 μm) at positions $(x,y)$ pass through the MPLC system, consisting of 7 phase plates separated by ~25mm of free-space propagation, implemented using a spatial light modulator and a mirror. Through these 7 planes each input spot at position $x,y$ is mapped to a corresponding Hermite-Gaussian mode $(m,n)$ (MFD=400μm), which is in turn transformed into the Laguerre-Gaussian basis through use of two cylindrical lenses.

[1] Nokia Bell Labs, 791 Holmdel Rd., Holmdel, NJ 07722, USA  [2] School of Information Technology and Electrical Engineering, The University of Queensland, Brisbane, QLD, 4072, Australia. [†]These authors contributed equally: Nicolas K. Fontaine, Joel Carpenter. *e-mail: j.carpenter@uq.edu.au

Approaches based on sorting $N$ modes through a cascade of $N$-1 interferometers[6], [10], [11] are inherently difficult to scale to large mode counts. Another practical consideration of any mode sorter is the spatial profile of the output beams on which the spatial components are mapped. Ideally, the sorted output spatial components are spatially separated Gaussian beams that are identical in terms of waist parameter. This enables diffraction-limited performance, compatibility with single-mode fibre for detectors and sources, and the efficient usage of a given optical system's available clear aperture and numerical aperture, assisting scalability to large mode counts. However, most current approaches natively output a beam which is non-Gaussian, not circularly symmetric, and with spot properties that differ between modes[7], [8], [10]–[12], [28], meaning additional mode dependent optics would be required for each modal component in order for lossless compatibility with single-mode detectors or sources.

In this work, we have discovered that an important special class of transformation, Cartesian points to Cartesian indices of HG modes, can be performed using remarkably few phase planes. Here, over 210 modes are demonstrated using just 7 planes, rather than the 420 planes that could be required in the general case of transformation of 210 modes in some other basis. This special transformation allows both the azimuthal and radial components of a multimoded beam to be easily mapped to an equally spaced Cartesian grid of identical Gaussian spots. Over 210 HG modes are demonstrated using a single spatial light modulator and a mirror, with three additional lenses used to convert the HG modes to LG. In the Supplementary Material we also present a 325 mode device. Not only do these mode multiplexers operate in the important bases of HG/LG, but they are also the largest mode count multiplexers demonstrated of any kind.

**The difficulty of spatial mode decomposition**

So why is spatial decomposition so difficult when compared to decomposing wavelength or polarisation? Why despite some 25 years of research on the measurement and generation of LG beams, has no device previously been able to perform a true two-dimensional decomposition? Lossless separation of wave components, whether wavelength, polarisation or space, are always performed through phase manipulation and interference. Hence the ease at which components of a beam can be spatially separated depends upon the ease at which a different phase can be applied to each component. For wavelength and polarisation this is straightforward. Wavelength-dependent phase shifts can be applied through optical path length differences, and polarisation dependent phase shifts can be applied through birefringence and/or reflections, and this spatial phase is transformed to spatial separation through propagation. Importantly, imparting spatial phase does not change the wavelength or the polarisation. In the spatial domain, not only is there no straightforward way in general to apply independent differential phase to spatial components which occupy almost identical space, but application of spatial phase changes the spatial components of the beam as it propagates. Unlike wavelength or polarisation, the input and output spatial bases must evolve through the transition, requiring in general a three-dimensional refractive index profile.

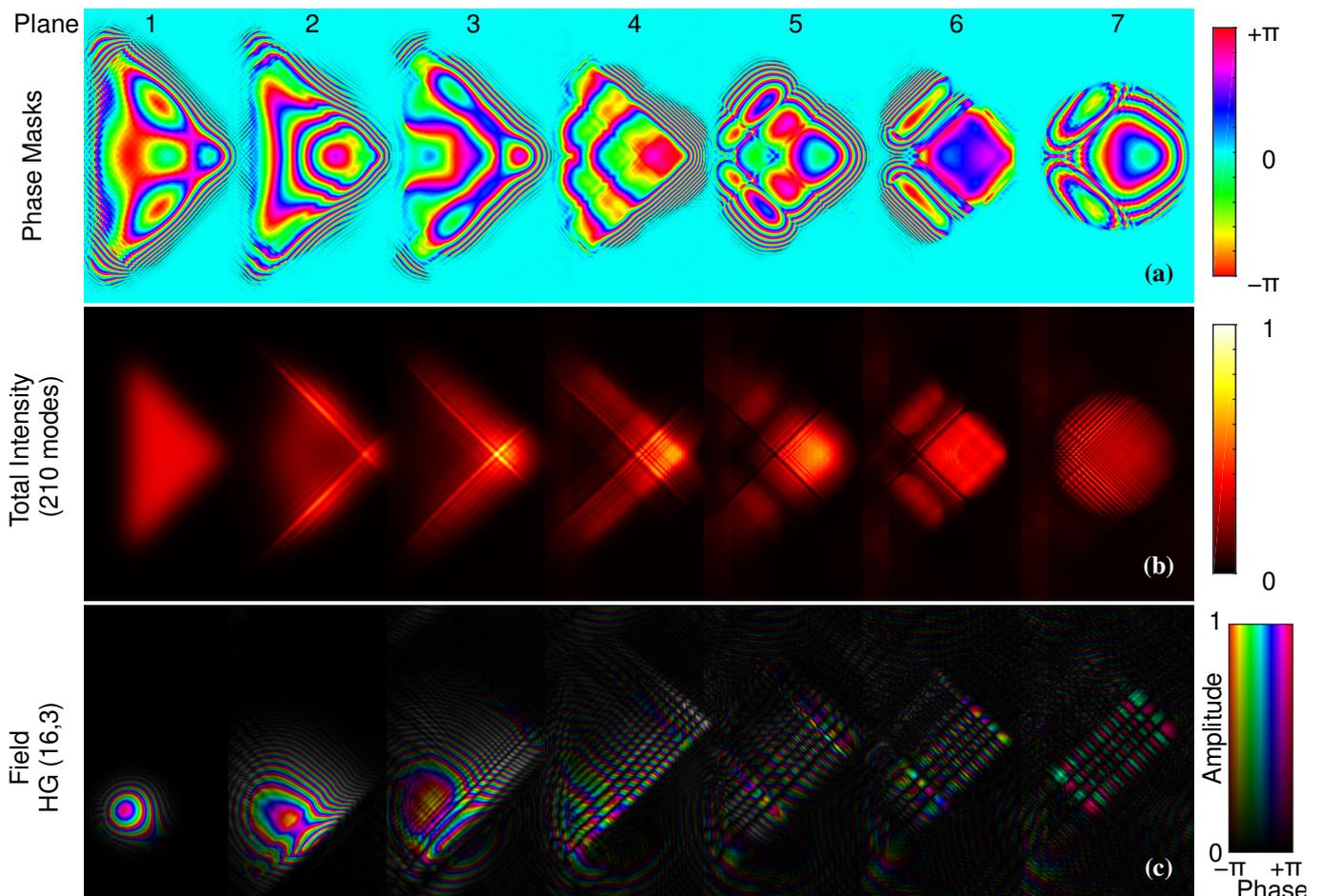

Figure 2 - Cartesian to Hermite-Gaussian transformation using multi-plane light conversion. (a) 7 phase planes used to perform the transformation. (b) Total intensity of the first 210 modes in each plane. (c) Example of the evolution of the complex amplitude of the HG16,3 mode through the device.

A photonic lantern[29], [30] is one example of a device which can decompose a beam into spatially separated components. It achieves this through an adiabatically tapered refractive index profile which transitions from a multimode fibre at one end, to an array of single-mode fibres at the other. However such a device is difficult to scale losslessly to large mode counts, particularly if the mapping of input modes to output spots must be controlled. A multi-plane light conversion system[31] (MPLC) comprises multiple phase plates separated by free-space propagation and in some ways emulates a discretised version of the continuous photonic lantern but is more amenable to implementing arbitrary spatial transformations. In the general case, the number of phase planes required scales as twice the total number of spatial modes, again making large mode counts difficult in practice. However not every kind of spatial basis transformation requires this full $2N$ phase planes, and some important modal decompositions and transformations require far less. Conversion between HG and LG requires only two cylindrical lenses[2], [32]–[34], and log-polar-based azimuthal mode-sorters[7], [8], [12] also require just two planes of phase manipulation regardless of the number of modes. Of course a lens was another example, performing a Fourier decomposition using a single plane of parabolic phase.

**Cartesian to HG to LG transformation**

Here we find another important basis transformation, HG to Cartesian can be easily performed using surprisingly few phase planes. While parabolic phase, which maps $x,y$ position to $x,y$ tilt is used to perform Fourier decomposition, cubic phase, which maps $x,y$ position to $x,y$ focus, is crucial for this transformation. The mapping itself is illustrated in Figure 1. Each Gaussian spot in a Cartesian array at position $(x,y)$ in Figure 1 is mapped to a single HG mode $(m,n)$ using an MPLC system[31], [35], [36]. The output HG basis can then be transformed to the LG basis if desired through cylindrical lenses[2], [37]. A schematic of the MPLC system which transforms HG modes to Cartesian coordinates (and vice-versa) is illustrated in Figure 1 and Figure 2, and consists of very few optical components. Namely, an equally spaced Cartesian array of single-mode fibres collimated by microlenses, a spatial light modulator (SLM) and a mirror. For conversion to the LG basis, two additional cylindrical lenses are required. Where appropriate, the fibre array could simply be a pixelated camera or some other detector array. Importantly, not only are the optical components standard items which can be purchased off-the-shelf without any custom fabrication, but the total number of optical components is almost independent of the number of modes being sorted. The transformation is performed using just seven planes of phase manipulation for over 210 modes. The Supplementary Material also demonstrates another example of 325 modes. The phase planes are separated by only free-space propagation and the folded design makes the device very compact. Figure 2a indicates that the phase masks are of low spatial resolution and can be described using relatively few pixels compared with the SLM itself. A summary of the transformation for the first 210 spatial modes is shown in Figure 2b and for a specific example of the $HG_{16,3}$ mode in Figure 2c. The masks are calculated using an inverse design process, known as wavefront matching[38] or adjoint optimisation[39]–[41]. The algorithm is surprisingly simple and effective and the optimisation process is visualised in the Supplementary Video. In short, the algorithm[36] attempts to match the phases of each pair of input and output modes at all points in space using the discrete phase planes. The masks for each of the seven planes are calculated numerically by propagating the desired basis (Cartesian grid of Gaussian spots) at one end through the optical system, and the corresponding desired output basis (HG modes) in the backwards direction. The phase masks are then updated iteratively until convergence by numerically propagating from plane-to-plane, backwards and forwards through the device. At each step, the phase mask is updated to become the phase of the superposition of the overlaps between each pair of input (A) and output modes (B). That is, the phase masks at each step become the average phase error between the modes propagating forward (A) and the modes propagating backwards (B). Specifically, the phase ($\phi$) at each plane at each step is given by, $\phi=\arg\{\sum_{i=1}^{N} A_i B_i^*\}$, where $N$ is the total number of modes (210 in this case), and $A_i$ and $B_i$ are the $i$-th modes in the forward and backward direction respectively. In a similar fashion to a Gerchberg-Saxton[42] type approach, the algorithm implements a steepest-descent search, but it is not guided by any error function. Rather it reaches convergence by continually enforcing phase matching at each iteration step.

Example commented Matlab code is included to demonstrate the procedure in the Supplementary Materials, as are pre-calculated phase mask sets, which can be used as-is, for several common SLM models currently on the market. These pre-calculated masks can be easily scaled to other dimensions and wavelengths without recalculation. The transformation is an approximation and is not strictly unique, although all low-loss solutions have similar features. The transformation is based largely on cubic phase manipulations, which generate Airy-like beams that are superimposed together to approximate Hermite-Gaussians. Some illustrative examples are detailed in the Supplementary Materials.

**Experimental results, 210 mode device**

The schematic of the MPLC device itself is shown in Figure1 and Figure 2, with the entire characterisation apparatus shown in Figure 3a. The device implemented here consists of an input array of Gaussian beams with mode-field diameter (MFD) of 60μm, and a square array pitch of $127/\sqrt{2}=89.8$μm. These spots propagate 20mm before the first reflection off the SLM, a Holoeye PLUTO-II with a dielectric backplane for high reflectivity (>95%). Light is then reflected back and forth between the SLM and a silver mirror parallel to the SLM 12.5mm away (~25mm propagation between planes), undergoing seven reflections off the SLM, before exiting the device as HG modes with MFD of 400μm. From there, a Fourier lens, $f$=160mm is used to focus the beam onto an InGaAs camera for characterisation in the HG basis, or through an additional pair of $f$=200mm cylindrical lenses to transform into the LG basis. Off-axis digital holography[1], [43] is performed to reconstruct the amplitude and phase of the output beam for each input spot in the array, for all modes over a wavelength range of 1510nm to 1620nm. As illustrated on the right of Figure 2a, digital holography simply measures the intensity of the interference between the mode being measured, (S), and a tilted reference quasi-plane wave, (R). This intensity, $|S+R|^2$, is then numerically Fourier transformed, the desired term selected, inverse Fourier transformed back into the plane of the original image, and the original tilt of the reference wave R removed. Yielding the recovered optical field, S. Although performed digitally, this is analogous to physically focusing the intensity pattern $|S+R|^2$ with a positive lens and picking off the desired part of the Fourier-transformed field containing information about S, with a pinhole aperture. The recovered field, S, can then be numerically overlapped with all Laguerre-Gaussian modes yielding the complex amplitude of each mode contained in that field. The output beam could have been analysed using any one of a number of modal decomposition techniques[44], [45]. The advantage of digital holography in this context is not only that it captures full amplitude and phase information regarding all modes, but that it does so by adding only minimal additional optics, minimising the effect of the measurement apparatus itself on the finally measured result. Once the field is recovered digitally, all optical alignment and mode generation on the LG/HG side of the MPLC device can be essentially perfect, as this is done numerically in post-processing from theoretical ideals, rather than having to physically implement a separate device for these operations which ultimately becomes part of the device-under-test being characterised[7], [8], [10]–[12].

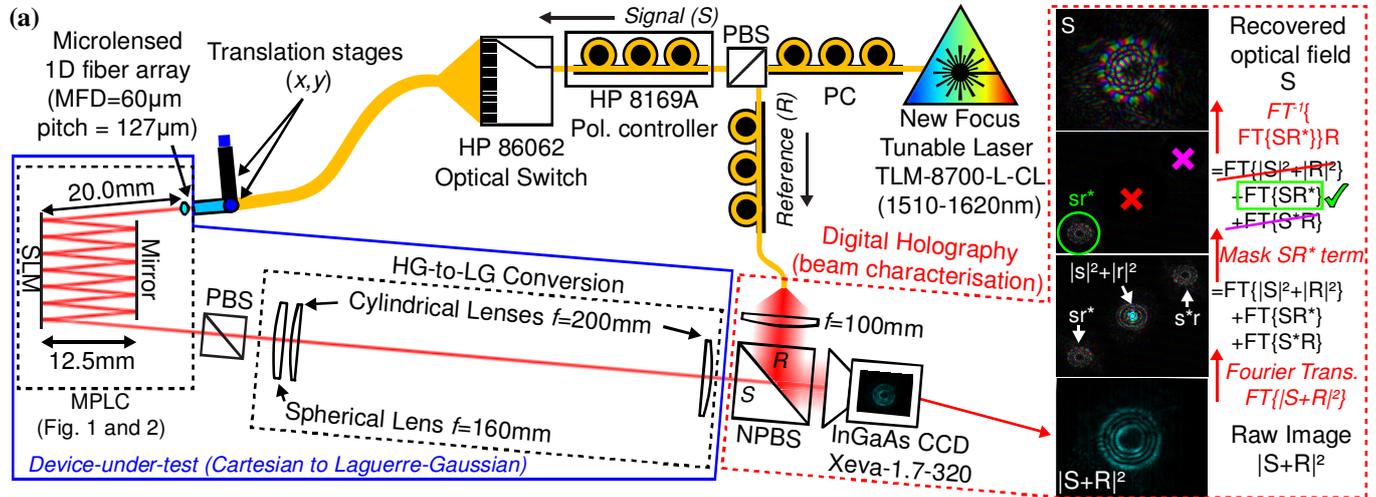

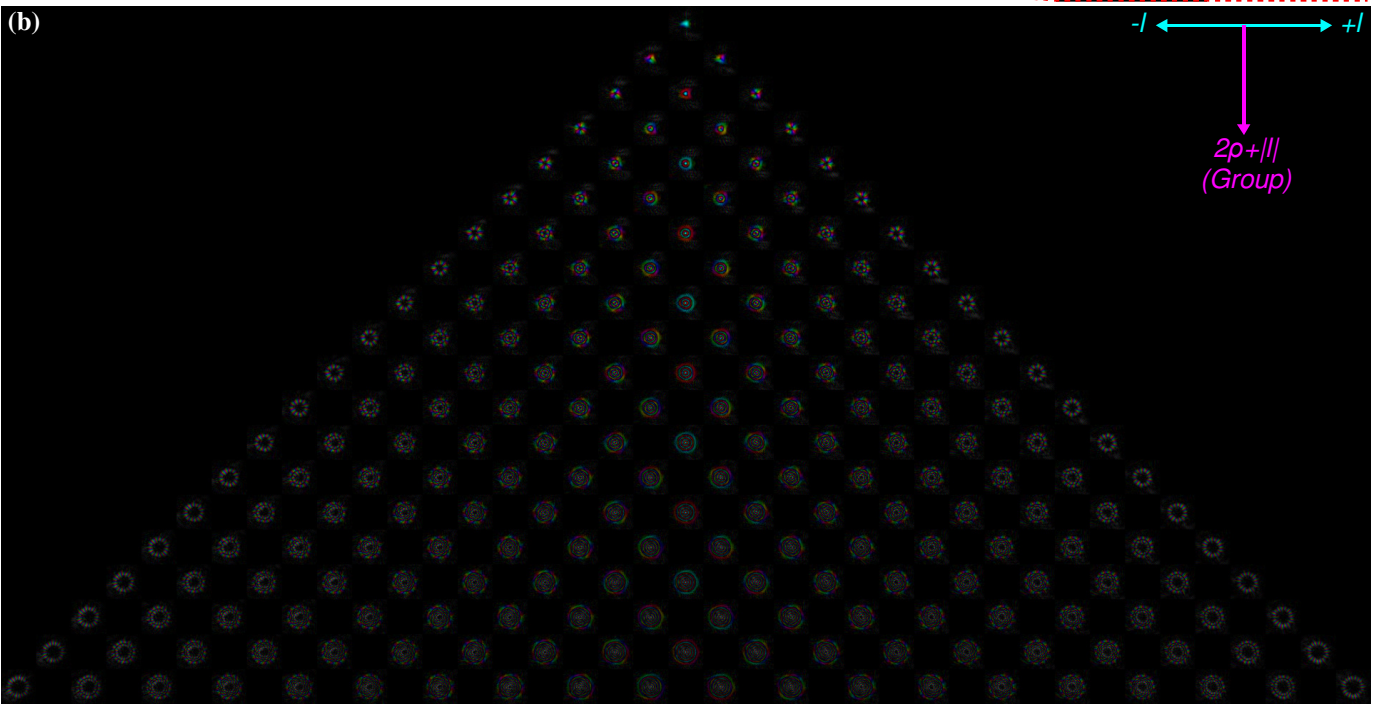

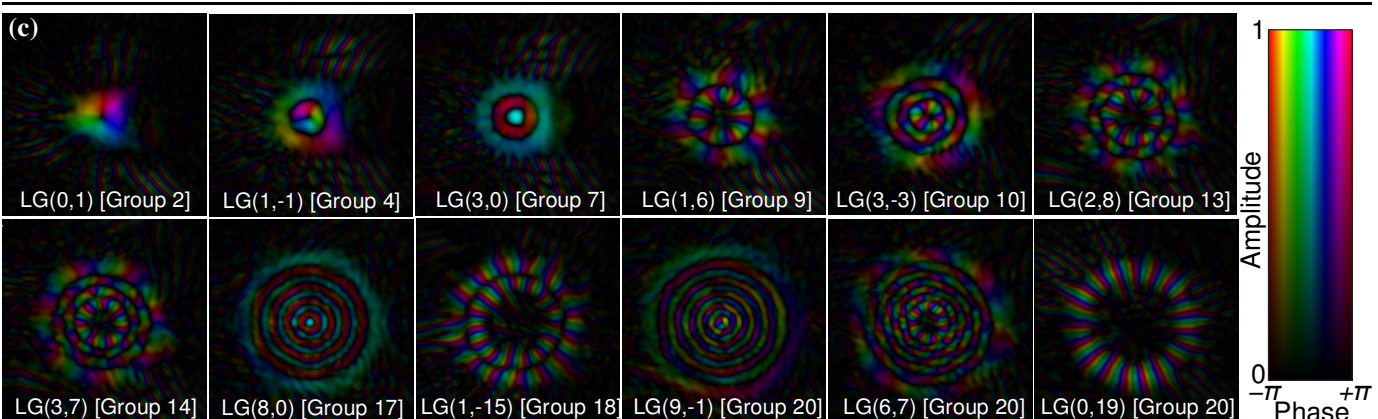

Figure 3 - Measured optical fields at 1565nm. (a) Measurement apparatus based on off-axis digital holography (b) Composite image of the full set of 210 modes. Azimuthal index runs left-to-right, mode-group runs top-to-bottom. Full-resolution image in Supplementary Material. (c) Select modes of various order.

With the full output optical field recovered for every input mode, the complete linear behaviour of the device is acquired. The MPLC device is now described by an $N \times N$ complex matrix which contains the amplitude and phase of the coupling between all pairs of input/outputs modes, as a function of wavelength[46]–[50]. From these matrices any linear property of the device can be extracted. The matrices for both the simulated and measured device are available in the Supplementary Material, from which the reader can calculate any linear parameter of interest. Additional detail on the experimental apparatus and procedure is available in the Supplementary Material.

A composite image of the full set of measured optical fields at the centre wavelength for all 210 modes (20 mode groups) is illustrated in Figure 3a. Figure 3b provides examples of various higher-order and lower-order modes of various radial and azimuthal indices and degenerate mode-group order. All modes have the correct number of rings ($\rho$) and helical phase ($l$) profiles. The full 210 mode set in full resolution is provided in the Supplementary Material, as are a HG example and a 325 mode example. The results are quantified in two different bases; using the singular value decomposition (SVD) of the transfer matrix, as well as in the device's native Laguerre-Gaussian basis. The SVD takes the transfer matrix of the device $T$ and expresses it as the product of three matrices, $T = U\Sigma V^*$. Where $U$ and $V$ are unitary transformations of the input basis (Gaussian spots) and output basis (HG/LG modes) such that $\Sigma$ is a real diagonal matrix containing the singular values. That is, the SVD finds the input and output basis through the device such that there is no crosstalk between input and output channels, only loss. The metrics the SVD yields are independent of the basis the device was originally characterised in. The singular values are especially relevant to coherent communications employing multiple-input, multiple-output processing (MIMO) as they are related to the channel capacity. The highest and lowest singular values represents the lowest and highest loss mode superposition through the device respectively. The ratio between these two extreme singular values is the condition number of the transfer matrix, and its square is the mode dependent loss (MDL), a measure of how 'invertible' the matrix is. An in-depth discussion of IL, MDL and the singular value decomposition (SVD) is provided in the Supplementary Material.

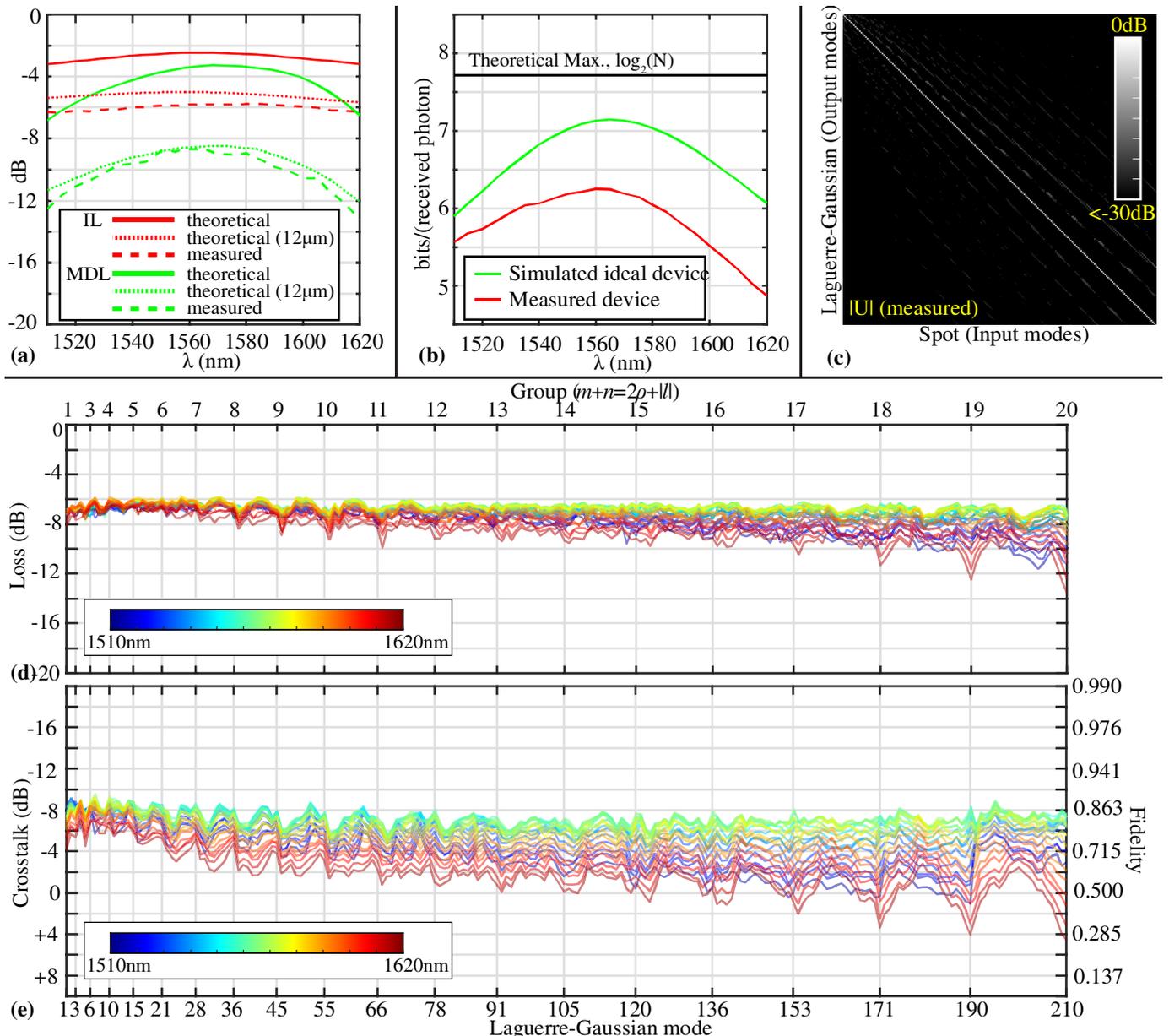

Figure 4 - Measured properties of LG mode sorter for 210 azimuthal and radial components (20 groups). (a) Simulated and measured insertion loss (IL) and mode-dependent loss (MDL). (b) Simulated and measured information capacity per received photon. (c) Measured amplitude of transfer matrix at centre wavelength (1565nm). (d) Measured loss of LG modes. (e) Measured total crosstalk of LG modes

The theoretical performance of the transformation is shown in Figure 4a. Insertion loss (IL) is defined as the average loss over all possible modal superpositions through the device (average squared singular value), and mode dependent loss (MDL) is the largest possible difference in loss between any two modal superpositions through the device (ratio between the largest and smallest singular value squared). Theoretically, for a lossless SLM and mirror, the transformation has an insertion loss of 2.5dB at the centre wavelength, increasing to 3.2dB at 1510nm and 1620nm. MDL is theoretically 3.3dB at the centre wavelength, 6.8dB at 1510nm and 6.5dB at 1620nm. Experimentally, the observed IL is between 5.8dB and 6.3dB, which for the centre wavelength corresponds with approximately 0.82dB of total loss, or 0.49dB of excess loss per reflection from the SLM. MDL was measured to be between 8.7dB at the centre wavelength, 12.5dB at 1510nm and 13.3dB at 1620nm. Which is a measured MDL comparable to mode multiplexers with an order of magnitude less spatial modes[21], or even measurements of near-unitary components such as short lengths of multimode fibre[49], [51]. It should be noted, that MDL is not the difference in loss between the maximum and minimum loss LG mode of the device. Those losses, the diagonal elements of the transfer matrix (Figure 4c) are illustrated in Figure 4d. The maximum variation in loss between any two LG modes is 1.7dB at the centre wavelength. Loss and crosstalk at the centre wavelength does not strongly depend on the order of the mode group, but the overall loss and crosstalk levels do depend on the total number of modes supported. Wavelength dependence does tend to get worse for higher-order modes as these modes contain higher spatial frequencies and must diffract over a larger path length from the edges of the Cartesian array. Similar principles apply when transformations are calculated to support increasing number of modes. As mode count increases, the performance of all modes tends to degrade together as a whole, but there is more degradation in bandwidth than there is in overall performance at the centre wavelength. Discussion and examples on this topic available in the Supplementary Material.

Inter-pixel crosstalk and the finite spatial resolution of the SLM[52], [53] is a significant contributing factor to MDL. Although the phase of the masks is smooth and of low spatial frequency, when implemented with finite phase stroke on the SLM, phase wraps are introduced, and these sudden discontinuities occur with finite spatial resolution due to the liquid crystal material properties. Example simulation results that convolve the phase level of each SLM pixel with a Gaussian of width 12μm are illustrated in Figure 4a for reference. Blurring of the phase levels causes scatter and crosstalk between modes, primarily into the next mode-group. Phase wrap related scatter would be less relevant if the masks were implemented using a lithographic process of higher spatial resolution, an SLM with sufficient phase stroke such that wraps could be reduced or avoided, or simply less modes were supported.

The measured transfer matrix of Figure 4c, yields a channel capacity, shown in Figure 4b of 6.25 bits/photon compared to the simulated ideal device of 7.15 bits/photon, or absolute theoretical maximum of $\log_2(210)$ = 7.71 bits/photon. The worst LG mode at the centre wavelength has a total crosstalk of -5.5dB, defined as the power in the desired mode relative to the total power in all other modes. Average total crosstalk over all 210 modes is -7.2dB. Crosstalk per mode is the above values divided by 210, yielding -28.7dB and -30.4dB for the worst-case and average respectively. Several additional plots and the measurement matrices as Matlab files are provided in the Supplementary Material.

We have demonstrated an MPLC-based mode-sorter capable of generating the first 210 modes in the LG basis using just an SLM and a mirror. This device can be easily implemented using common optical components and allows the spatial properties of light to be decomposed in 2D and with high dimensionality. Enabling functionality in the spatial domain, which is already common in the spectral and polarisation domains.

**References**

[1] B. E. A. Saleh and M. C. Teich, *Fundamentals of Photonics*, 2nd ed. New York, USA: John Wiley & Sons, Inc., 1991.

[2] L. Allen, M. W. Beijersbergen, R. J. C. Spreeuw, and J. P. Woerdman, "Orbital angular momentum of light and the transformation of Laguerre–Gaussian laser modes," *Phys. Rev. A*, vol. 45, pp. 8185–8189, 1992.

[3] A. Mair, A. Vaziri, G. Weihs, and A. Zeilinger, "Entanglement of the orbital angular momentum states of photons," *Nature*, vol. 412, p. 313, Jul. 2001.

[4] W. N. Plick and M. Krenn, "Physical meaning of the radial index of Laguerre-Gauss beams," *Phys. Rev. A*, vol. 92, no. 6, p. 63841, Dec. 2015.

[5] E. Karimi and E. Santamato, "Radial coherent and intelligent states of paraxial wave equation," *Opt. Lett.*, vol. 37, no. 13, pp. 2484–2486, 2012.

[6] J. Leach, J. Courtial, K. Skeldon, S. M. Barnett, S. Franke-Arnold, and M. J. Padgett, "Interferometric Methods to Measure Orbital and Spin, or the Total Angular Momentum of a Single Photon," *Phys. Rev. Lett.*, vol. 92, no. 1, p. 13601, Jan. 2004.

[7] G. C. G. Berkhout, M. P. J. Lavery, J. Courtial, M. W. Beijersbergen, and M. J. Padgett, "Efficient Sorting of Orbital Angular Momentum States of Light," *Phys. Rev. Lett.*, vol. 105, no. 15, p. 153601, Oct. 2010.

[8] M. Mirhosseini, M. Malik, Z. Shi, and R. W. Boyd, "Efficient separation of the orbital angular momentum eigenstates of light," *Nat. Commun.*, vol. 4, p. 2781, Nov. 2013.

[9] W. N. Plick, R. Lapkiewicz, S. Ramelow, and A. Zeilinger, "The Forgotten Quantum Number: A short note on the radial modes of Laguerre-Gauss beams," *arXiv Prepr. arXiv1306.6517*.

[10] Y. Zhou *et al.*, "Sorting Photons by Radial Quantum Number," *Phys. Rev. Lett.*, vol. 119, no. 26, p. 263602, Dec. 2017.

[11] X. Gu, M. Krenn, M. Erhard, and A. Zeilinger, "Gouy Phase Radial Mode Sorter for Light: Concepts and Experiments," *Phys. Rev. Lett.*, vol. 120, no. 10, p. 103601, Mar. 2018.

[12] A. Dudley *et al.*, "Efficient sorting of Bessel beams," *Opt. Express*, vol. 21, no. 1, pp. 165–171, 2013.

[13] K. Wagner *et al.*, "Entangling the Spatial Properties of Laser Beams," *Science (80-. ).*, vol. 321, no. 5888, p. 541 LP-543, Jul. 2008.

[14] R. Fickler *et al.*, "Quantum Entanglement of High Angular Momenta," *Science (80-. ).*, vol. 338, no. 6107, p. 640 LP-643, Nov. 2012.

[15] A. C. Dada, J. Leach, G. S. Buller, M. J. Padgett, and E. Andersson, "Experimental high-dimensional two-photon entanglement and violations of generalized Bell inequalities," *Nat. Phys.*, vol. 7, p. 677, May 2011.

[16] M. M. and O. S. M.-L. and M. N. O. and B. R. and M. M. and M. P. J. L. and M. J. P. and D. J. G. and R. W. Boyd, "High-dimensional quantum cryptography with twisted light," *New J. Phys.*, vol. 17, no. 3, p. 33033, 2015.

[17] M. Krenn, M. Malik, M. Erhard, and A. Zeilinger, "Orbital angular momentum of photons and the entanglement of Laguerre–Gaussian modes," *Philos. Trans. R. Soc. A Math. Phys. Eng. Sci.*, vol. 375, no. 2087, Feb. 2017.

[18] Y. Zhang *et al.*, "Simultaneous entanglement swapping of multiple orbital angular momentum states of light," *Nat. Commun.*, vol. 8, no. 1, p. 632, 2017.

[19] A. Sit *et al.*, "High-dimensional intracity quantum cryptography with structured photons," *Optica*, vol. 4, no. 9, pp. 1006–1010, 2017.

[20] E. Karimi *et al.*, "Exploring the quantum nature of the radial degree of freedom of a photon via Hong-Ou-Mandel interference," *Phys. Rev. A*, vol. 89, no. 1, p. 13829, Jan. 2014.

[21] N. K. Fontaine *et al.*, "30× 30 MIMO transmission over 15 spatial modes," in *Optical Fiber Communication Conference*, 2015, p. Th5C–1.

[22] J. Wang *et al.*, "Terabit free-space data transmission employing orbital angular momentum multiplexing," *Nat. Photonics*, vol. 6, p. 488, Jun. 2012.

[23] N. Zhao, X. Li, G. Li, and J. M. Kahn, "Capacity limits of spatially multiplexed free-space communication," *Nat. Photonics*, vol. 9, p. 822, Nov. 2015.

[24] D. J. Richardson, J. M. Fini, and L. E. Nelson, "Space-division multiplexing in optical fibres," *Nat. Photonics*, vol. 7, no. 5, pp. 354–362, 2013.

[25] R. Ryf *et al.*, "Mode-Division Multiplexing Over 96 km of Few-Mode Fiber Using Coherent 6x6 MIMO Processing," *J. Light. Technol.*, vol. 30, no. 4, pp. 521–531, 2012.

[26] M. K. and R. F. and M. F. and J. H. and M. M. and T. S. and R. U. and A. Zeilinger, "Communication with spatially modulated light through turbulent air across Vienna," *New J. Phys.*, vol. 16, no. 11, p. 113028, 2014.

[27] G. Gibson *et al.*, "Free-space information transfer using light beams carrying orbital angular momentum," *Opt. Express*, vol. 12, no. 22, pp. 5448–5456, 2004.

[28] M. P. J. L. and D. J. R. and A. S. and J. C. and N. K. S. and G. A. T. and A. E. W. and M. J. Padgett, "Efficient measurement of an optical orbital-angular-momentum spectrum comprising more than 50 states," *New J. Phys.*, vol. 15, no. 1, p. 13024, 2013.

[29] S. G. Leon-Saval, A. Argyros, and J. Bland-Hawthorn, "Photonic lanterns," *Nanophotonics*, vol. 2. p. 429, 2013.

[30] S. G. Leon-Saval, N. K. Fontaine, J. R. Salazar-Gil, B. Ercan, R. Ryf, and J. Bland-Hawthorn, "Mode-selective photonic lanterns for space-division multiplexing," *Opt. Express*, vol. 22, no. 1, pp. 1036–1044, 2014.

[31] J.-F. Morizur *et al.*, "Programmable unitary spatial mode manipulation," *J. Opt. Soc. Am. A*, vol. 27, no. 11, pp. 2524–2531, 2010.

[32] E. Abramochkin and V. Volostnikov, "Beam transformations and nontransformed beams," *Opt. Commun.*, vol. 83, no. 1, pp. 123–135, 1991.

[33] M. W. Beijersbergen, L. Allen, H. E. L. O. van der Veen, and J. P. Woerdman, "Astigmatic laser mode converters and transfer of orbital angular momentum," *Opt. Commun.*, vol. 96, no. 1, pp. 123–132, 1993.

[34] E. G. A. and V. G. Volostnikov, "Generalized Gaussian beams," *J. Opt. A Pure Appl. Opt.*, vol. 6, no. 5, p. S157, 2004.

[35] G. Labroille, B. Denolle, P. Jian, P. Genevaux, N. Treps, and J.-F. Morizur, "Efficient and mode selective spatial mode multiplexer based on multi-plane light conversion," *Opt. Express*, vol. 22, no. 13, pp. 15599–15607, 2014.

[36] N. K. Fontaine, R. Ryf, H. Chen, N. David, and J. Carpenter, "Design of High Order Mode-Multiplexers using Multiplane Light Conversion," in *European Conference on Optical Communication (ECOC)*, 2017, p. Tu.1.F.4.

[37] J. Courtial and M. J. Padgett, "Performance of a cylindrical lens mode converter for producing Laguerre–Gaussian laser modes," *Opt. Commun.*, vol. 159, no. 1, pp. 13–18, 1999.

[38] Y. Sakamaki, T. Saida, T. Hashimoto, and H. Takahashi, "New Optical Waveguide Design Based on Wavefront Matching Method," *J. Light. Technol.*, vol. 25, no. 11, pp. 3511–3518, 2007.

[39] A. Y. Piggott, J. Lu, K. G. Lagoudakis, J. Petykiewicz, T. M. Babinec, and J. Vučković, "Inverse design and demonstration of a compact and broadband on-chip wavelength demultiplexer," *Nat. Photonics*, vol. 9, p. 374, May 2015.

[40] C. M. Lalau-Keraly, S. Bhargava, O. D. Miller, and E. Yablonovitch, "Adjoint shape optimization applied to electromagnetic design," *Opt. Express*, vol. 21, no. 18, pp. 21693–21701, 2013.

[41] J. Lu and J. Vučković, "Nanophotonic computational design," *Opt. Express*, vol. 21, no. 11, pp. 13351–13367, 2013.

[42] G. Yang, B. Dong, B. Gu, J. Zhuang, and O. K. Ersoy, "Gerchberg–Saxton and Yang–Gu algorithms for phase retrieval in a nonunitary transform system: a comparison," *Appl. Opt.*, vol. 33, no. 2, pp. 209–218, 1994.

[43] J. W. Goodman and R. W. Lawrence, "Digital image formation from electronically detected holograms," *Appl. Phys. Lett.*, vol. 11, no. 3, pp. 77–79, Aug. 1967.

[44] A. Forbes, A. Dudley, and M. McLaren, "Creation and detection of optical modes with spatial light modulators," *Adv. Opt. Photonics*, vol. 8, no. 2, pp. 200–227, 2016.

[45] T. Kaiser, D. Flamm, S. Schröter, and M. Duparré, "Complete modal decomposition for optical fibers using CGH-based correlation filters," *Opt. Express*, vol. 17, no. 11, p. 9347, 2009.

[46] J. Carpenter, B. J. Eggleton, and J. Schröder, "Observation of Eisenbud-Wigner-Smith states as principal modes in multimode fibre," *Nat. Photonics*, vol. 9, no. 11, 2015.

[47] N. K. Fontaine *et al.*, "Characterization of Space-Division Multiplexing Systems using a Swept-Wavelength Interferometer," in *Optical Fiber Communication Conference/National Fiber Optic Engineers Conference 2013*, 2013, p. OW1K.2.

[48] J. Carpenter, B. J. B. J. Eggleton, and J. Schröder, "Complete spatiotemporal characterization and optical transfer matrix inversion of a 420 mode fiber," *Opt. Lett.*, vol. 41, no. 23, pp. 5580–5583, 2016.

[49] J. Carpenter, B. J. B. J. Eggleton, and J. Schröder, "Comparison of principal modes and spatial eigenmodes in multimode optical fibre," *Laser Photon. Rev.*, vol. 11, no. 1, p. 1600259–n/a, Dec. 2016.

[50] J. Carpenter, B. J. B. J. Eggleton, and J. Schröder, "Reconfigurable spatially-diverse optical vector network analyser," *Opt. Express*, vol. 22, no. 3, pp. 2706–2713, 2014.

[51] J. Carpenter, B. J. Eggleton, and J. Schröder, "Observation of Eisenbud–Wigner–Smith states as principal modes in multimode fibre," *Nat Phot.*, vol. 9, no. 11, pp. 751–757, Nov. 2015.

[52] T. Lu, M. Pivnenko, B. Robertson, and D. Chu, "Pixel-level fringing-effect model to describe the phase profile and diffraction efficiency of a liquid crystal on silicon device," *Appl. Opt.*, vol. 54, no. 19, pp. 5903–5910, 2015.

[53] B. Apter, U. Efron, and E. Bahat-Treidel, "On the fringing-field effect in liquid-crystal beam-steering devices," *Appl. Opt.*, vol. 43, no. 1, pp. 11–19, 2004.